\documentclass[conference]{IEEEtran}

\usepackage{booktabs,multirow}
\usepackage{pict2e}
\usepackage{diagbox}

\ifCLASSINFOpdf
  \usepackage[pdftex]{graphicx}
\else
  \usepackage[dvips]{graphicx}
\fi
\begin{document}
\title{DeepEmotex: Classifying Emotion in Text Messages using Deep Transfer Learning} 

%
%
%

\author{\IEEEauthorblockN{Maryam Hasan}
\IEEEauthorblockA{Department of Computer Science\\
Worcester Polytechnic Institute\\
and San Francisco State University\\
Email: mhasan@wpi.edu}
\and
\IEEEauthorblockN{Elke Rundensteiner}
\IEEEauthorblockA{Department of Computer Science\\ and Data Science\\
Worcester Polytechnic Institute\\
Email: rundenst@wpi.edu}
\and
\IEEEauthorblockN{Emmanuel Agu}
\IEEEauthorblockA{Department of Computer Science\\
Worcester Polytechnic Institute\\
Email: agu@wpi.edu}}

\maketitle


\begin{abstract}
Transfer learning 
has been widely used in natural language processing through deep pretrained language models, such as Bidirectional Encoder Representations from Transformers and Universal Sentence Encoder.   
Despite the great success,  language models  get overfitted when applied to small datasets and are prone to  forgetting when fine-tuned with a classifier.   
To remedy this problem of forgetting in transferring deep pretrained language models from one domain to another domain, existing efforts explore fine-tuning methods to forget less. 
We propose DeepEmotex an effective  sequential transfer  learning method to detect emotion in text. 
To avoid forgetting problem, the fine-tuning step is instrumented by a large amount of emotion-labeled data collected from Twitter.  

We conduct an experimental study using both curated Twitter data sets and benchmark data sets.  
DeepEmotex models  achieve over  91\%  accuracy for multi-class emotion classification on test dataset. 
We evaluate the performance of the fine-tuned DeepEmotex models in classifying emotion in  EmoInt  and Stimulus benchmark datasets. 
The models correctly classify emotion in 73\% of the instances in the benchmark datasets. 
The proposed DeepEmotex-BERT model outperforms Bi-LSTM result on the benchmark datasets by 23\%. We also study  the effect of the size of the fine-tuning dataset on  the  accuracy  of  our  models. Our evaluation results show that fine-tuning with a large set of emotion-labeled data improves both the robustness and effectiveness of the resulting target task model. 


  
\end{abstract}


%
\IEEEpeerreviewmaketitle

\section{Introduction}

Detecting emotions in text is a challenging problem because of the semantic ambiguity of the emotion expression in text and fuzzy boundaries of emotion classes \cite{hasan2019automatic}.  Moreover, the context can completely change the emotion for a sentence as compared to perceived emotion when the sentence is evaluated standalone.  For example, the sentence ``I started crying when I realized!''  will be perceived as a sad feeling, however considering it in the context ``I just qualified for the scholarship. I started crying when I realized!'',  it turns out to be a happy emotion.  Similarly, the sentence ``My cat gets angry when I work from home!'' within the context ``My cat gets angry when I work from home. It is funny!''.
Another example is the sentence ``Try to do that again!''  is very likely to be perceived as no-emotion, however a majority will judge it as an Angry feeling with the context ``How dare you make fun of me like that!  Try to do that again!''.  Therefore, understanding the context is important to detect emotion in text.

Feature selection and representation are important tasks for emotion classification in texts due to the high dimensionality of text features and the existence of irrelevant features.  The performance of classification algorithms strongly depends on feature selection and representation \cite{bengio2013representation}. 

Supervised learning methods have 
proven to be promising in emotion classification. However, these methods are domain dependent, which means that a model built on one domain (e.g., messages on a
specific topic or event) 
may perform poorly on another domain. 
The reason is that rather distinct words and phrases domain-specific words may be used to express emotion in different domains.  Table \ref{keywords} shows common keywords for two different domains, death of George Floyd and Covid-19 pandemic. 
For example,  keywords \textit{justice, protest, violent, murder, and racism} are domain-specific words characteristic of the first domain whereas  keywords such as \textit{death, disease, caugh, and fever}  are Covid19-specific keywords. 
Due to the mismatch of common keywords between these different domains, an emotion classifier trained on one domain may not work well when directly applied to another domain. Therefore, cross-domain emotion classification
methods are desirable. Such methods reduce domain dependency and the costs and human labor required for manually labeling a sufficient number of example texts required for training supervised learning models by transferring knowledge from related  domains to the domain of interest  \cite{pan2009survey}.

For decades, supervised learning methods solving NLP problems have been based on shallow models (e.g., SVM and logistic regression) trained on very high dimensional, sparse  and hand-crafted features  \cite{young2018recent}. In the last few years, neural networks based on a rich variety of 
representations have been producing 
much superior to prior winning results on various NLP tasks  
\cite{young2018recent}. This trend is sparked by the success of distributed word embeddings including Word2Vec and GloVe   \cite{mikolov2013distributed, pennington2014glove}. 

\begin{table}%
\caption{Example keywords of different domains}
\centering
{
\begin{tabular}{ |p{2.87cm}| p{4.4cm}|}
\hline    
 Death of George Floyd & attack, violence, protest, curfew, death,   murder, racism, black lives \\
 \hline
 Covid-19 pandemic & fever, death, patient, disease, vaccine, mask, immunity, pandemic \\
\hline
\end{tabular}}
\label{keywords}
\end{table}

Deep neural network-based approaches have achieved remarkable successes on text classification tasks 
by being pretrained on huge volumes of text archives such as Wikipedia. 
Deep learning models automatically learn multiple layers of low dimensional feature representations 
and thus reduce the need for hand-crafted features \cite{young2018recent}. 
Deep learning methods have been utilized in learning word representations 
through neural language models \cite{mikolov2013distributed}. 
Along with the success of deep learning in many other application domains, deep learning has also become popular for solving  sentiment and emotion classification in recent years \cite{young2018recent}. 

Different neural network architectures have emerged to deal with challenges in NLP tasks, such as RNNs, LSTM, BiLSTM. More recently, the
Transformers architecture \cite {vaswani2017attention} 
has been shown to have extremely promising results for NLP tasks.  
Transformers language models have shown
better language understanding abilities that allow them to achieve state-of-the-art results for many different NLP tasks. 

While deep learning models have achieved
state-of-the-art results on text classification tasks, these models are trained from scratch, requiring huge input datasets, and days to converge \cite{howard2018universal}. 
Instead of learning from scratch, transfer learning 
can be used to transfer knowledge from a general-purpose domain and task into a more specialized target domain and task \cite{raffel2020exploring, houlsby2019parameter, ruder2019transfer}. 
In fact, transfer learning in many cases can achieve or even exceed the performance of traditional deep learning models trained for the particular task from scratch, yet while only requiring a much smaller set of labeled examples for the fine-tuning to the target task  \cite{howard2018universal}. Transfer learning can also be beneficial for sentiment and emotion classification in text. 

Transfer learning 
has been widely used in natural language processing (NLP) through pretrained Language Models (LMs) 
\cite{goodfellow2016deep}. 
Deep pretrained Language Models, such as  Universal Sentence Encoder (USE)  \cite{cer2018universal} 
and Bidirectional Encoder Representations from Transformers (BERT) \cite{devlin2018bert}, have been widely used in Natural Language Processing (NLP). 
Despite the great success, LMs get overfitted to small datasets and are prone to catastrophic forgetting when fine-tuned with a classifier  \cite{howard2018universal}. 
 To remedy the catastrophic forgetting in transferring deep pretrained LMs, existing efforts mainly explore fine-tuning tricks to forget less. 
Fine-tuning experiments rely on the pretrained language model parameters. 
 During target task training, the
language model must adapt enough to be able to solve the target task involving a  different input distribution and output label space than the pretrained model. The adaptation step must also avoid overfitting or forgetting of what was learned during pretraining \cite{phang2018sentence}. 

We propose DeepEmotex, an effective sequential transfer learning method to detect emotion in text.  
DeepEmotex is an extension of our previous work called Emotex \cite{emotex1}. Emotex requires extensive hand-crafted features to achieve a high performance. 
Such hand-crafted features are time-consuming to create and they are often incomplete. 
To solve this problem, we develop a deep transfer learning approach 
to be able to extract domain-specific features based on the context instead of using static hand-crafted features. 
 
To avoid overfitting to our target task, DeepEmotex utilizes a large set of emotion-labeled data easily obtainable in Twitter. We study the effects of varying the amount of data used for fine-tuning the pre-trained models. 
We verify if increasing the amount of fine-tuning data reduces the effect of the forgetting problem in transferring knowledge from the pre-tranied language domain  (i.e., BERT) to our target domain. 
Our evaluation results show that fine-tuning with a large set of emotion-labeled data improves both the robustness and effectiveness of the resulting target task model. 
Overall, we made the following contributions in this paper. 

\begin{itemize}
    \item 
    We develop DeepEmotex to classify emotion in text messages using deep transfer learning. 
    DeepEmotex develops transfer learning  models for fine-tuning pre-trained language models and  learns emotion specific features which are more contextually aware of a new domain. 
    
    \item 
    DeepEmotex utilizes two state-of-the-art pre-trained models, known as BERT \cite{devlin2018bert} and USE  \cite{cer2018universal}. 
    To avoid overfitting to the target task we utilize a large set of  emotion-labeled data collected from Twitter messages. 

    \item 
    We analyze the adaptation or fine-tuning phase during which the
    pretrained knowledge is transferred to our emotion classification task. 
    DeepEmotex achieves 92\% classification accuracy on our test data by fine-tuning BERT. 
    
    \item
    We evaluate the performance of DeepEmotex models to classify emotion in  EmoInt and Stimulus benchmark datasets. DeepEmotex models were able to correctly classify emotion in 73\% of  the instances in the benchmark datasets. 

    \item
     We  verify the effect of the size of data for  the fine-tuning task on  the  accuracy  of  our  models. Our evaluation results show that fine-tuning with a large set (more than 30K) of emotion-labeled data improves both the robustness and effectiveness of the resulting target task model. 
    
    \item
    We evaluate DeepEmotex models by comparing their results with a baseline model (i.e., Bi-LSTM). 
    Our evaluation results show that the proposed DeepEmotex-BERT model outperforms  the baseline  model  by  23\%. 

\end{itemize}




The remainder of the paper is organized as follows: we first overview related literature on emotion classification in text using deep learning methods in Section \ref{related}. 
A brief background knowledge in sequential transfer learning is provided in Section \ref{transfer}. 
Section \ref{deepEmotex} describes the details of the proposed DeepEmotex model to detect emotion using transfer learning. 
Section \ref{experiment}  outlines the experimental setup, and Section \ref{results} discusses the empirical results and analysis. Finally, Section \ref{conclude} presents the conclusion and introduces the next research direction.



\section{Related Work}
\label{related}
Recently, approaches which employ deep learning methods for emotion and sentiment detection in text have been proposed. Deep learning based methods use distributed word embeddings as input, which already encode some semantic and syntactic information. 

Ren. et al. \cite{ren2016context}  proposed a context based neural network model for Twitter sentiment analysis 
by incorporating contextualized features from relevant Tweets into the model in the form of word embedding vectors. 
They showed that improvements can be achieved by modeling the context of a given target tweet using neural pooling functions to extract the most useful features from tweets. 
Another context-sensitive method for sentiment classification proposed by Teng et al. \cite{teng2016context}. Their method is based on a simple weighted-sum model, using bidirectional LSTM to learn the sentiment strength, intensification and negation of lexicon sentiments in composing the sentiment value of a sentence.

Wang et al. \cite{wang2016dimensional} combined CNN and LSTM to capture 
both local information within sentences and long-distance dependency across sentences. They proposed a regional CNN-LSTM model, which consists of two parts: regional CNN and LSTM, to predict the valence and arousal ratings of text. 

Due to the lack of emotion-labelled datasets, many supervised classification algorithms for emotions have been done on data gathered from microblog such as Twitter, using hashtags or emoticons as the emotion label for the data. 

Felbo et al. \cite{felbo2017using}  used millions of emoji occurrences in social networks as noisy labels for pre-training neural models in order to  learn representations of emotional contexts. 
To capture the context of each word they used two bidirectional LSTM layers with 1024 hidden units  (512 in each direction), with an attention layer that takes all of LSTM layers as input using skip-connections. 
Their results confirmed that the distant supervision to a more diverse set of noisy labels enables the models to learn better representations of emotional content in text and  obtain better performance for detecting sentiment, emotions and sarcasm. 

Yu et al. \cite{yu2018improving} proposed a transfer learning method using LSTM and a dual attention network.
They divided the sentence representation into two feature spaces to capture the general sentiment words and emotion-specific words. 
A similar approach \cite{gupta2017sentiment} utilizes a custom LSTM architecture in order to assign emotion labels to conversations in social media. 
They labelled their dataset through a
heuristic procedure and then reconstructed this heuristic with their classifier. 

 Koper et al. \cite{koper2017ims} predicted  emotion intensity in tweets by applying deep learning method with extended lexicons of affective norms. 
They showed that domain-specific embeddings (trained on twitter data) perform superior to other embeddings. 

Polignano et al.  \cite{polignano2019comparison} 
studied that an architecture
composed of BiLSTM and self-attention
demonstrated promising results in different
datasets. 
the authors used transfer learning, using word embeddings already pre-trained in different domains. Thus, they could reduce the computational cost while covering a wider range of terms regardless of domains. 
The performance of their  model for the ISEAR dataset is about \%62. 

Sent2affect proposed by Kratzwald et al.  \cite{kratzwald2018deep} is a  form of transfer learning for affective computing.  Their BiLSTM  network is pre-trained for a different task (i.e.  sentiment analysis), while the output layer is subsequently tuned to the task of emotion recognition.  The
resulting performance is evaluated across 6 benchmark datasets, 
They found BiLSTM with pre-trained word embeddings as
the superior method in all experiments.  We further identify that the BiLSTM appears to outperform the unidirectional LSTM in all experiments. 

Batbaatar et al. \cite{batbaatar2019semantic} 
proposed a semantic-emotion neural network (SENN) that combines two sub-networks to capture semantic and emotional information. The first network consists of a BiLSTM to capture semantic information.  
The second network is a CNN to capture
emotional information. 
The performance of the SENN model for the ISEAR dataset, is about \%74. 
Imran et al. \cite{imran2020cross} analyzed emotion of citizens from different cultures to the  Coronavirus. LSTM models used for estimating the sentiment polarity and emotions from extracted tweets have been trained to achieve state-of-the-art accuracy on the sentiment140 dataset. 

Deep learning based methods mostly use  distributed word vectors.
Commonly used embeddings are Word2Vec \cite{mikolov2013distributed},
GloVe \cite{pennington2014glove}, and FastText \cite{bojanowski2017enriching}. 
Word2Vec and GloVe treat words as the smallest  units. FastText uses a different approach where it treats each
individual word as being made of n-gram characters. 
Thus, it can handle rare words which are not present in the dictionary. 

More recently, contextualized word embeddings are proposed, called 
USE \cite{cer2018universal} and BERT  \cite{devlin2018bert}, to incorporate context information 
in conventional word embeddings. 
However, these word embeddings are generalized on various tasks and limited to provide emotion
information, therefore learning task-specific emotion embedding with the neural network has been proven to be effective. 


Some researchers predict emotion in textual dialogues. 
Luo and Wang \cite{luo2019emotionx} fine-tune BERT model to predict emotion in dialogues  by  choosing  from  four  emotion  classes,  joy,  sadness,  anger, and neutral. They use datasets consisting of scripts from the TV show, Friends, and the anonymous Facebook chat logs named EmotionPush. 
Chatterjee et al.  \cite{chatterjee2019semeval} use  BiLSTM model to infer the underlying emotion from textual dialogues by choosing from four emotion classes, happy, sad, angry, and other. 
EmoDet2 is another work to classify emotion in dialogues developed by Al-Omari et al. \cite{al2020emodet2}. They classify EmoContext dataset into happy, sad, angry and other. They use GloVe  embeddings and features extracted from  AffectiveTweets. They also extract word contextual embeddings from BERT model. These vectors feed feed-forward and BiLSTM models to obtain predictions. Their result show that the performance of the system  is increased by extracting BERT embeddings then feeding them into an BiLSTM network. 


\section{Sequential Transfer Learning}
\label{transfer}
Transfer learning aims to leverage knowledge from a source task to improve the performance of a model in a different, but related target task. 
Sequential transfer learning learns source tasks and target tasks in sequence, and transfers knowledge from source tasks to improve the models  performance on target tasks \cite{raffel2020exploring, ruder2019transfer}. 

Sequential transfer learning typically consists of two stages: pretraining
and adaptation. During pretraining, the model is trained on source tasks. During adaptation, the pretrained model is
further trained on target tasks. The standard adaptation methods includes fine-tuning and feature extraction. Fine-tuning updates parameters of the pretrained model, while feature extraction considers the pretrained model as a feature extractor and keeps the parameters fixed during the adaptation phase  \cite{felbo2017using,chen2020recall}. 

Sequential transfer learning has been widely
used recently, and deep pretrained language models have achieved great success on various NLP tasks \cite{devlin2018bert, peters2018deep}. While the adaptation of the deep pretrained models  is very efficient, it is prone to forgetting, where the model forgets previously learned knowledge from source tasks when learning target tasks \cite{howard2018universal}. 

\section{DeepEmotex: Emotion Detection using  Sequential Transfer Learning}
\label{deepEmotex}

Deep pretrained language models, such as USE  \cite{cer2018universal} and BERT \cite{devlin2018bert}, have been used in natural language processing.  A wide range of NLP tasks has been promoted by 
these pretrained language models 
through Sequential Transfer
Learning \cite{howard2018universal}: pretrain a language model
on large-scale unlabeled data and then adapt it to
downstream tasks. 
The adaptation step is usually
conducted in two methods: fine-tuning or freezing pretrained weights. In practice, fine-tuning is 
adopted more widely due to its flexibility 
\cite{peters2019tune, lan2019albert}.

DeepEmotex utilizes sequential transfer learning approach, 
where source and target tasks are learned in sequence.  
That means, models are not optimized jointly as in multi-task learning but each task is learned separately  \cite{ruder2019transfer}. 

Sequential transfer learning  consists  of  two  stages: a pre-training phase in which general-purpose  representations are learned on a source task or domain, and an adaptation or fine-tuning 
phase during which the learned knowledge is transferred to a target task or domain \cite{ruder2019transfer}. 

Using deep neural networks architectures is not trivial for emotion classification. 
The problem is that the pretrained models are not suited to the small emotion datasets. They  typically get overfitted 
and suffer  forgetting problem when fine-tuned with a classifier  \cite{howard2018universal}.
This is especially a problem when the fine-tuning data  is  small. 
To remedy the  forgetting problem in transferring deep pretrained models, existing efforts mainly explore fine-tuning tricks to forget less. 
To fine-tune with less forgetting, we 
utilize a large set of emotion-labeled data easily obtainable in Twitter. 

Most prior work has focused on different pretraining objectives to learn general-purpose word or sentence representations  \cite{mikolov2013distributed, kiros2015skip}.  
A few works have explored the fine-tuning phase 
and how to adapt the pretrained model to a given target task \cite{peters2019tune,chen2020recall,phang2018sentence, ragheb2019attention}. 
There are two common approaches for fine-tuning: 
The first approach is to use the pre-training network  as a feature extractor \cite{donahue2014decaf}, where all layers in the model are frozen when fine-tuning on the target task except the last layer. 
In this approach the pre-trained representations are used in a downstream model. 
Alternatively, the pre-trained model's parameters are  unfrozen and fine-tuned on a new task  \cite{dai2015semi}. 
This approach enables to adapt a general-purpose representation to many different tasks. 

\begin{figure}[!t]
\centering
\includegraphics[width=\linewidth,height=66mm]{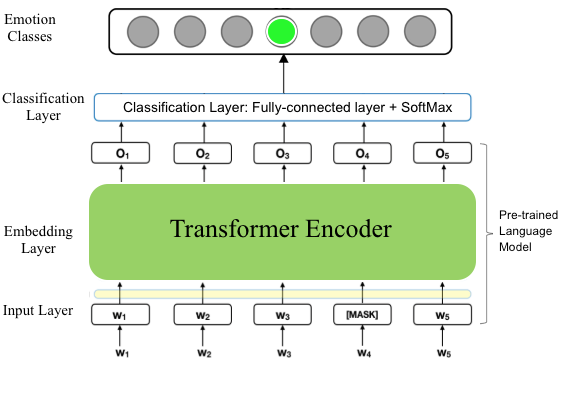}
\caption{Model of DeepEmotex. 
The embedding layer 
learns an embedding that contains the semantic textual information in input text. 
The learned representations are fed into the classification
layer for emotion prediction.} 
\label{deepemotex}
\end{figure}

Figure \ref{deepemotex} shows our DeepEmotex model. 
Our model represents input words by their embeddings. Following the embedding layer, our model consists of a transformer encoder, 
followed by a SoftMax classification layer. 
Gaining a better understanding of the adaptation phase is key in making the most use out
of pre-trained representations.  Accordingly, 
DeepEmotex utilizes two state-of-the-art pre-trained models,  
known as BERT \cite{devlin2018bert} and Universal Sentence Encoder \cite{cer2018universal}.  
Using these models, we transfer knowledge learned from a large corpus to our emotion classification model.   
We then fine-tune DeepEmotex model to fit 
to our target emotion datasets. 


\subsection{DeepEmotex-USE: A Transfer Learning Model using Universal Sentence Encoder}

Universal Sentence Encoder (USE) \cite{cer2018universal} is a  deep neural network to create universal sentence embeddings. 
Universal embeddings are  pretrained embeddings obtained from training deep learning models on a huge corpus. These pretrained (generic) embeddings can be used in a wide variety of NLP tasks including text classification, semantic similarity and clustering. 

The Universal Sentence Encoder is trained and optimized for greater-than-word length text, such as sentences, phrases or short paragraphs.  
The input is a variable length text. 
The Universal Sentence Encoder encodes the input text into 512-dimensional embeddings. 
The USE embeddings are trained on different data sources and tasks with the aim of dynamically accommodating a wide variety of natural language understanding tasks which require modeling the meaning of word sequences rather than just individual words. 


Essentially, there are  two versions of the USE models. The first version makes use of 
a Deep Averaging Network (DAN) where input embeddings for words and bi-grams are first averaged together and then passed through a feed-forward deep neural network (DNN) to produce sentence embeddings \cite{cer2018universal}. 
Deep Averaging Network (DAN) is simpler than the second version. 
The primary advantage of the DAN encoder is 
that its compute time is linear in the length of the input sequence. 

The second version makes use of 
the transformer-network based sentence encoding model. 
The transformer encoder is composed of a stack of N = 6 identical layers. Each layer has two sub-layers. The first is a multi-head self-attention mechanism, and the second is a simple, position-wise fully connected feed-forward network, followed by layer normalization. 

Their results \cite{cer2018universal} 
demonstrate that the transformer-based encoder achieves
the best overall transfer task performance. However,
this comes at the cost of computing time and
memory usage scaling dramatically with sentence
length. 
For our emotion classification task we use the transformer-based encoder as it achieves better overall performance than the DAN encoder. 


We apply transfer learning leveraging prior knowledge from pre-trained embeddings to solve our emotion classification task. 
Our sentence embedding sub-network  leverages the Universal Sentence Encoder. 
We fine-tune the sentence embeddings using our collected emotion-labeled dataset. 

We build a 
feed-forward neural network with two hidden dense layers and the rectified linear activation function (ReLU). 
 ReLU overcomes the vanishing gradient problem.  
This is good for deep neural networks which suffer from the vanishing and explosion gradient problem   
\cite{zhang2018deep}. 

A dense layer provides learning features from all the combinations of the features of the previous layer.
The input of our model is 512-feature vectors created using the Universal Sentence Encoder technology. The resulting vector is then fed into fully connected layers culminating in a softmax layer. 
We then fine-tune our model using collected labeled tweets introduced in Section \ref{deepdata}. We fine-tune the embedding weights by setting the trainable parameter to true. Here we leverage  transfer learning in the form of pre-trained embeddings. 

\begin{figure}[ht!]
\centering
\includegraphics[width=\linewidth]{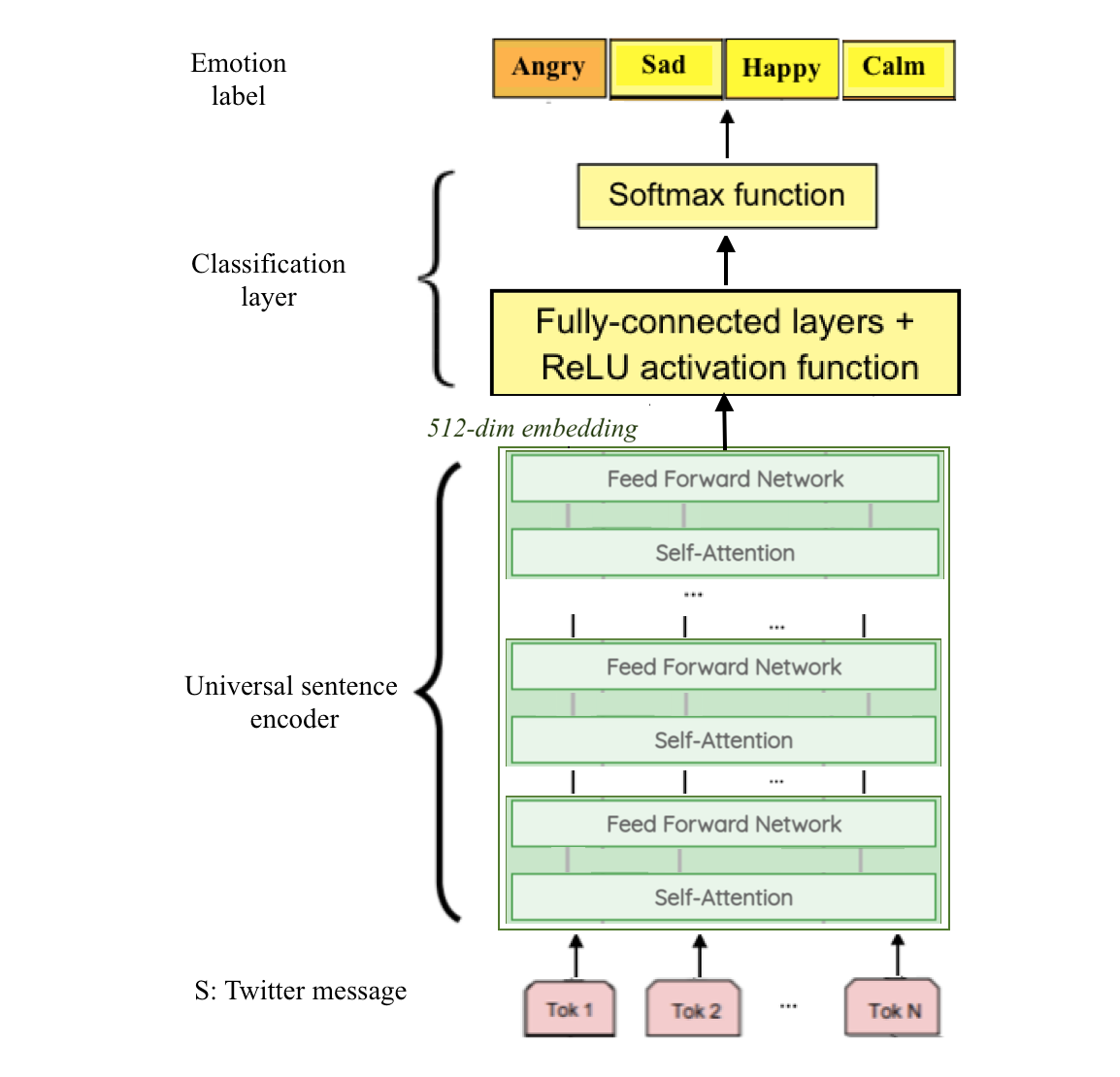}
\caption{Model of DeepEmotex-USE to classify emotion in text messages using Universal Sentence Encoder.} 
\label{use}
\end{figure} 

The overall architecture of DeepEmotex-USE is shown in Figure \ref{use}. The input of the model is a twitter message. 
First, the embedding layer uses the pre-trained  USE model to map a sentence into its embedding vector. The model that we are using 
splits the sentence into tokens, embeds each token and then combines them into context-aware 512-dimension embeddings. 
Then, the embeddings are passed  through a feed-forward  neural network with ReLU activation. It projects the input into 256-dimension embeddings and feeds  them  to the classification layer  to produce a classification probability. 
The output of our model is an emotion classification label.
The main objective is to correctly predict the emotion of each tweet.

\subsection{DeepEmotex-BERT: A Transfer Learning Model using Bidirectional Encoder Representations from Transformers }

\begin{figure}[ht!]
\centering
\includegraphics[width=\linewidth]{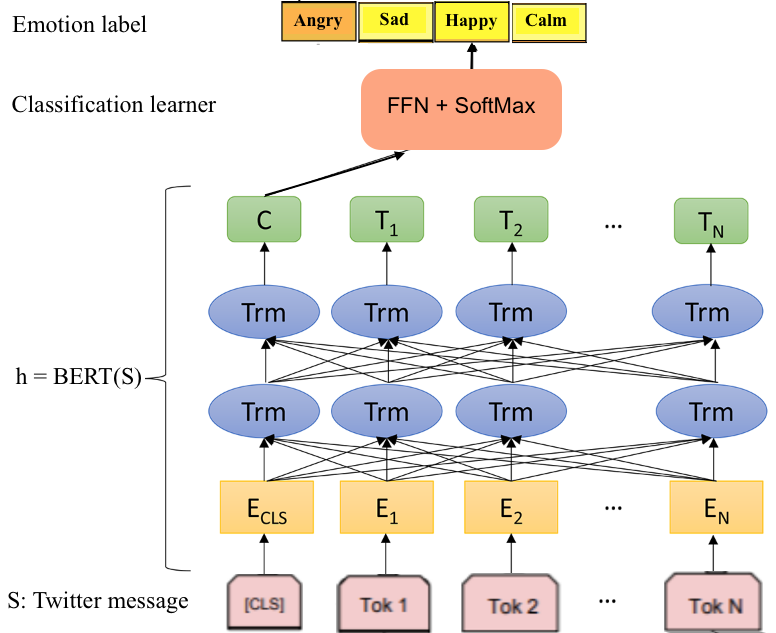}
\caption{Model of DeepEmotex to classify emotion in text messages using pretrained BERT.}
\label{bert-base}
\end{figure}

The  model architecture of DeepEmotex-BERT using Bidirectional Encoder Representations from Transformers (BERT) is shown in Figure \ref{bert-base}. The input of the model is a twitter message and the output is an emotion label. 
We use the pre-trained BERT model \cite{devlin2018bert}  to
generate  text representations. 
BERT learns text representations using a bidirectional Transformer encoder \cite{vaswani2017attention} pre-trained on the language modeling task. 
Transformers have a sequence-to-sequence model architecture. 
Each transformer includes a separate encoder and decoder component. The difference is in their use of attention known as self-attention. 
The core architecture consists of a stack of encoders fully connected to a stack of decoders. Each encoder consists of  a self-attention component and a feed forward network. Each decoder consists of a self-attention component, an encoder-decoder attention component, and a feed forward component. 

BERT has several variants based on model configurations. 
We adopt BERT-base \cite{vaswani2017attention} as our   base model. BERT-base consists of an encoder with 12-layer Transformer blocks. For each block in the encoder, it contains a 12-head self-attention layer and 768-dimensional hidden layer, yielding a total of 110M parameters. The base  model allows inputs up to a sequence of 512 tokens and outputs the vector representations of the sequence. The input sequence has one or two segments.  The first token
of the sequence is always $[CLS]$ which contains the special classification embedding. Another
special token $[SEP]$ is used for separating segments. In order to
facilitate the training and fine-tuning of BERT model, we transform the input text 
into $[CLS] + text + [SEP]$ format. 


We follow Devlin et al.  \cite{devlin2018bert} and create a fully connected layer over the final hidden state corresponding to the $[CLS]$ input token.
During fine-tuning, we optimize the
entire model, with the additional softmax classifier parameters $W \in R ^{K \times H}$, where $H$ 
is the dimension of the hidden state vectors and $K=4$ 
is the number of emotion classes. 

Let $S = ([CLS], t_1, \dots , t_m, [SEP], ..., t_n)$ be the input sequence (i.e., Twitter message), where $t_1, \dots, t_m$  denotes a sentence with m tokens. 
For emotion classification task, 
we use the final hidden state $h = BERT(S)$ 
of the first token $[CLS]$ 
as the representation of the whole sequence $S$. A standard softmax classifier added on top of BERT 
predicts the probability of emotion label c: 

$$x = W \cdot h + b $$
$$P(c|x) = softmax(x) = \frac {\exp (x) }   {\sum_{k=1}^{C} \exp (x) }$$

where $W$ is the task-specific weight matrix, and $b$ is the bias vector to be estimated. 
We fine-tune all parameters as well
as $W$ jointly by maximizing the log-probability of
the correct label. 

\section{DeepEmotex: Experimental Methodology}
\label{experiment}
After developing DeepEmotex models, we run several experiments to fine-tune pretranied  models based on the emotion-classification task using our input dataset.  
We conduct the experiments with the
deep pretrained language model BERT-base \cite{devlin2018bert}  and USE \cite{cer2018universal}. 

\subsection{DeepEmotex Emotion Dataset}
\label{deepdata}
Deep pretrained models get overfitted to small datasets and suffer  forgetting problem when fine-tuned with a classifier  \cite{howard2018universal}. 
To fine-tune with less forgetting, we 
utilize a large set of emotion-labeled data  obtainable in Twitter. 

In order to fine-tune pretranied models for our  emotion classification task, we need a large dataset of labeled tweets. Since there are not many human-labeled
datasets publicly available, we collect tweets with  emotion-carrying hashtags as a surrogate for emotion labels \cite{hasanusing, emotex1, mohammad2018semeval}. 
We define four emotion classes (i.e., joy, relax, anger, and  sadness)  based on the emotion model proposed by Circumplex model \cite{citeulike:630522}. 
We only collect the tweets labeled with these emotions. 

In order to collect enough tweets to serve as our labeled dataset, we develop a list of hashtags representing each of four  emotion classes proposed by the Circumplex model. For each emotion class, we define a  set of hashtags representing the emotion. 
We then use the set of hashtags to collect  tweets with the emotion hashtags at the end.  

One advantage of using hashtags for labeling emotion data is that the label is assigned by the writer of the tweet 
rather than an annotator who could wrongly decide the category of a tweet. After all, emotion is a fuzzy concept. Another advantage of this method is obviously that it enables us to acquire a sufficiently large labeled dataset to apply fine-tuning in our transfer learning approach. 

Twitter data is very noisy, not only because of
use of non-standard typography, but due to many duplicate tweets and the fact that tweets often have multiple
emotion hashtags. Since these reduce our ability
to build accurate models, we clean the data. 
We first apply some general data cleaning strategies. 
We convert the tweets to lower-case letters. We remove  non-ascii letters, urls,  
``@name'' and duplicate letters. 
We also filter out 
 all retweets based on existence of the
token ``RT'' regardless of its case.  
Since our goal is to create non-overlapping classes at the level of a tweet, we also remove 
all tweets with hashtags belonging to more than
one emotion of the four emotion classes. 

Table \ref{tweets} represents the number of labeled tweets selected for each class after pre-processing.  
The labeled data are relatively balanced. The number of data labeled relax is especially less than other classes. 
Our dataset comprises a total of 540,525 tweets labeled with four emotion classes. 
 We  shuffle our labeled datasets and create train, validation and test datasets.  
We train our models on a total of 300,000 tweets as our training dataset, validate on 60,000 tweets. We use 180,525 tweets as our test dataset. 
 
\begin{table}
\caption{Number of tweets collected as emotion-labeled data}
\centering
{
\begin{tabular}{|p{0.8cm} | p{0.9cm} |p{1cm}|p{1.1cm}|p{1.1cm}|p{0.8cm}|}
\hline  
Class & Happy-Active & Happy-Inactive & Unhappy-Active & Unhappy-Inactive & Total\\ \hline          

\#Tweets & 148,571  & 195,313 & 149,287 & 47,354 & 540,525 \\ \hline
\end{tabular}
}
\label{tweets}
\end{table}

\subsection{Experimental Setup of the DeepEmotex-USE  Model}
Universal Sentence Encoder model can be fine-tuned to our  target task in several ways by freezing layers to disable parameters updates. 
One common approach is to use the network as a feature extractor \cite{donahue2014decaf}, where all layers in the model are frozen during fine-tuning on the target task except the last layer. Another common approach
is to use the pre-trained model as an initialization, and thereafter the full model is
unfrozen \cite{felbo2017using}.  
We implement both approaches to fine-tune Universal Sentence Encoder. 
Our first transfer learning approach is implemented using the following workflow: 
\begin{itemize}
    \item Instantiate a base model and load pre-trained weights into it.
    \item Freeze all layers in the base model by setting trainable = False.
    \item 
    Add classification layer (Fully connected layer + SoftMax) on top of the frozen layers. 
    \item Train the new layers on our new dataset.  
\end{itemize}


Next, we implement the second approach. For this, we unfreeze all layers and re-train the whole model on our dataset for several epochs. 
This helps fine-tune the model towards our task by incrementally adapting the pre-trained features to our new data. 
We fine-tune the embedding weights by setting trainable=true. Weights are updated (via gradient descent) to minimize the loss during training. By training all layers of the model we are able to adjust the parameters 
across the network during back propagation. 
Our second transfer learning approach is implemented using the following workflow: 
\begin{itemize}
    \item Instantiate a base model and load pre-trained weights into it.
    \item Unfreeze all layers in the base model by setting trainable = True.
    \item 
    Add classification layer (Fully connected layer + SoftMax) on top of the USE unfrozen layers. 
    \item Train all the layers on our new dataset to fine-tune the old parameters on the new dataset. 
\end{itemize}

We use the Universal Sentence Encoder Version $3$ \footnote{https://tfhub.dev/google/universal-sentence-encoder-large/3} as our base model.  We add a feed-forward neural network with two hidden layers and the Relu activation function. 

We  set our batch size to 150 and number of epochs to 20. 
Batch-size defines the number of examples will be passed to our model during one iteration, and number of epochs is the number of times our model will go through the entire training set. 
We train the model on our training datasets using Adam optimizer with a learning rate of 0.001.  The performance of the re-trained model is evaluated at the end of training epochs with our test datasets. 



\subsection{Experimental Setup of the DeepEmotex-BERT Model}

BERT 
uses books corpus and Wikipedia 
data sources to pre-train 
their models. 
Although BERT aims to learn contextualized representations across a wide range
of NLP tasks, leveraging 
BERT alone still leaves the domain challenges unresolved as BERT is only trained on formal texts and has almost no understanding of social media text \cite{xu2019bert}.
The end tasks from the original BERT paper typically use tens of thousands of examples to ensure that the system is task-aware \cite{xu2019bert}. 
Below, we introduce  fine-tuning BERT to
boost the performance of classifying emotion on Twitter messages. 

We adopt the pre-trained BERT-BASE 
as the encoding layer of our DeepEmotex model. 
We extend them with extra tasks-specific layers and fine-tune the model on our emotion classification task. 
We focus on fine-tuning a classification layer implemented as a standard feed-forward and a 
softmax layer on top of the pre-trained BERT. 


For fine-tuning the target model, we keep 
the hyper-parameters the same as in the pre-training by Devlin  \cite{devlin2018bert},  
except for the batch size, learning rate, and number of training epochs. 
Devlin et al.  found the following range of possible values to work well across different tasks \cite{devlin2018bert}:
\begin{itemize}
    \item Batch size: 32, 64
    \item Learning rate: 5e-5, 3e-5, 2e-5
    \item Number of epochs: 2, 3, 4
\end{itemize}

The optimal hyper-parameter values
are task-specific. 
For our model, we decide to choose 
 a small learning rate, 
and train with a few epochs. 
Most of transfer learning models suffer from the so-called catastrophic forgetting problem. 
That is, the learnt knowledge is diminished when learning the new information from the target domain. 
Using a large learning rate such as 6e-4 makes the training fail to converge. 
The BERT authors \cite{devlin2018bert} recommend a number of training epochs between 2 and 4. Selecting a large number of epochs may cause over-fitting. 
As shown in Table \ref{bert-param}, 
we fine-tune our model for 2 epochs with a 
learning rate of 4e-5. 
A larger batch size often results in lower accuracy but faster epochs. 
In order to find the optimum batch size we perform several runs of varying batch sizes while keeping other parameters constant. 


\section{Experimental Results of DeepEmotex mdels}
\label{results}
In this section, we evaluate our approach and report empirical results. 

\subsection{Experimental Results of DeepEmotex-USE}
\label{results1}
We train our USE models (with  frozen and unfrozen layers) for 20 epochs. 
The training is conducted on our collected tweets. 
Our dataset comprises a total of 540,525 tweets labeled with four emotion classes as described in Section \ref{deepdata}. 
 We first shuffle our labeled datasets and create train, validation and test datasets.  
 The training data is used for fine-tuning and the testing dataset is used for evaluation. 

\begin{figure}[ht!]
\centering
\includegraphics[width=\linewidth,height=48mm]{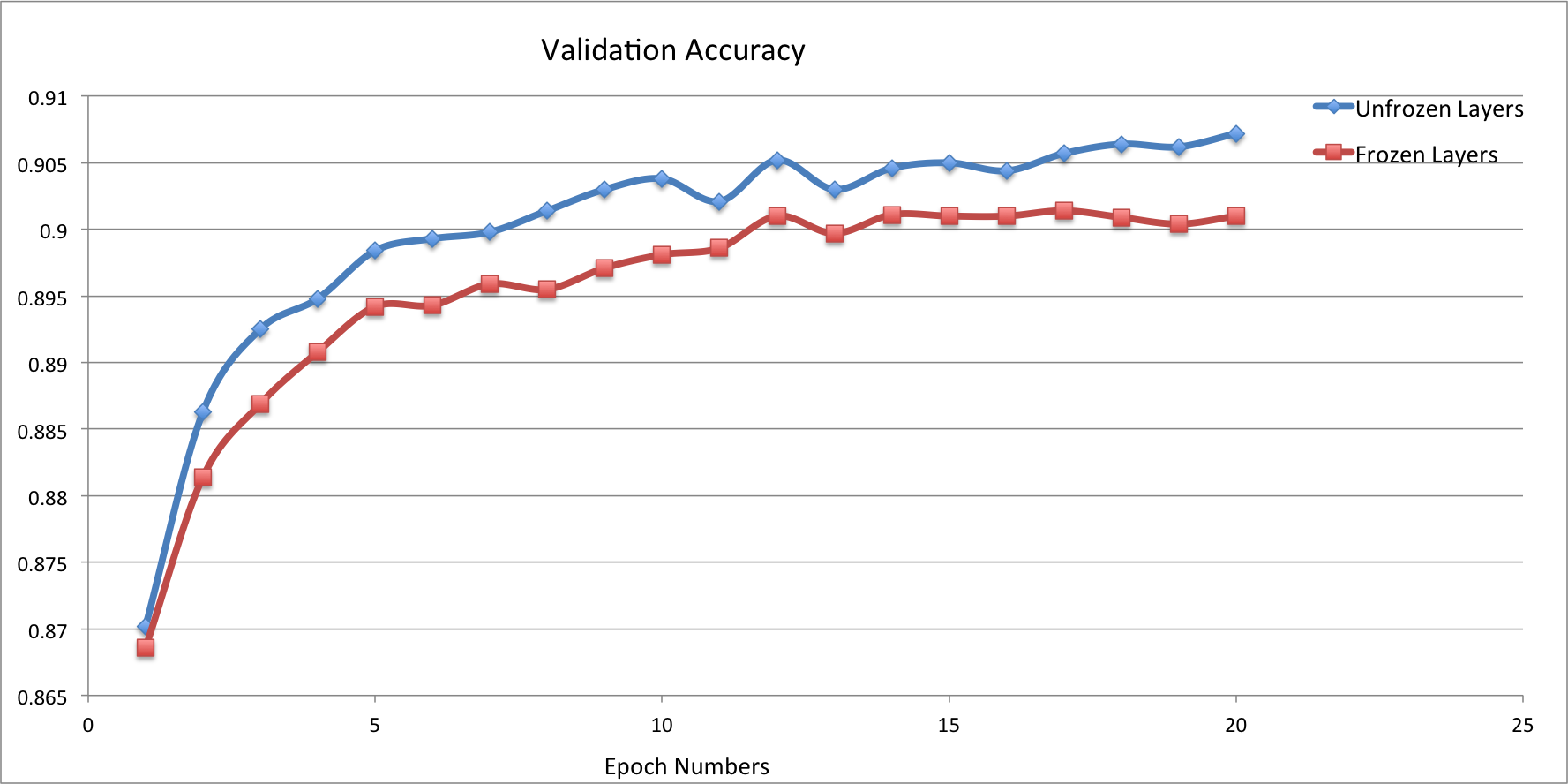}
\caption{Emotion Classification Accuracy of DeepEmotex-USE with frozen and unfrozen layers on validation set in terms of micro-average F1 score. } 
\label{use-epoch}
\end{figure}

Figure \ref{use-epoch} shows the classification accuracy of our two models with frozen and unfrozen layers on the validation set in terms of micro-average F1 score. 
As it shows, the accuracy of our models stabilize  at about epoch 14. 
The final validation results show that our model  with frozen layers gets an accuracy of $90\%\pm0.82\%$ (Mean $\pm$ Standard deviation) after training for 20 epochs.
The frozen model gets an accuracy of 90.6\% on our test dataset.  
This is consistent with our validation dataset. 

The final validation results show that our model with unfrozen layers gets an  accuracy of $90.6\%\pm0.85\%$ (Mean $\pm$ Standard deviation)  after training for 20 epochs. 
We get an accuracy of 91\% on our test dataset using the model with unfrozen layers.   
Comparing the results of the two models with frozen and unfrozen layers using statistical t-test 
shows that the unfreezing approach performed 
 better than the freezing approach ($p-value=0.007$). 

\subsection{Experimental Results of DeepEmotex-BERT}
\label{results2}

\begin{table}%
\caption{Parameters of fine-tuning BERT}
\centering
{
\begin{tabular}{|p{1.2cm} | p{0.7cm} |p{3cm}|p{1cm}|p{1cm}|}
\hline  
Parameters & Epoch & Batch sizes & Learning rate  \\ \hline  
Values  & 2 & 50,100,150,200,250 & 4e-5  \\ \hline          
\end{tabular}
} 
\label{bert-param}
\end{table}
Fine-tuning our BERT model is conducted on collected tweets. 
Our dataset comprises a total of 540,525 tweets labeled with four emotion classes. 
The collected tweets are shuffled and divided into training, validation and test datasets  as described in Section \ref{deepdata}. 

\begin{figure}[ht!]
\centering
\includegraphics[width=\linewidth]{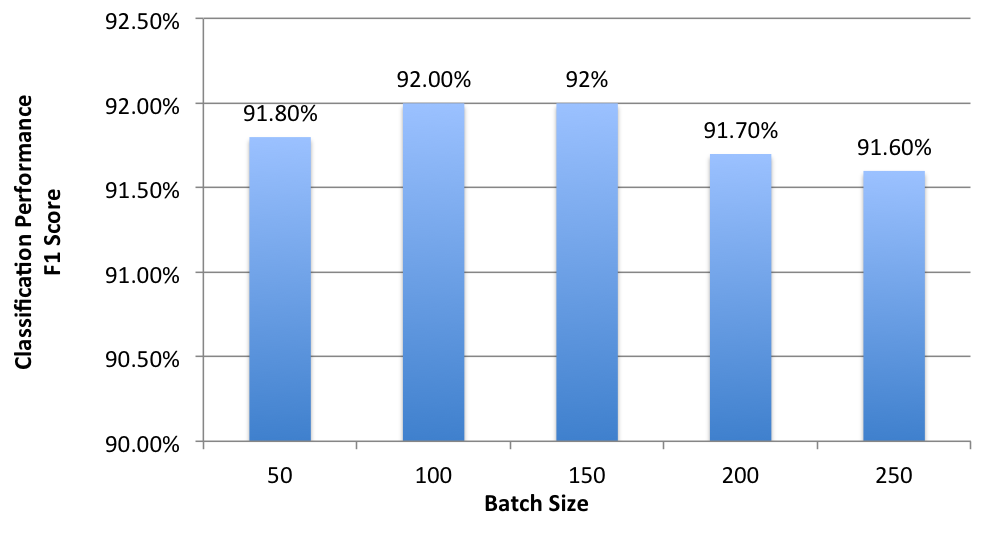}
\caption{Classification results of 
DeepEmotex-BERT 
on our test dataset using  different batch sizes in terms of micro-average F1 score.}
\label{bert-result}
\end{figure}

We fine-tune our BERT model  using  different  batch sizes for 2 epochs. 
  Figure \ref{bert-result} shows the classification performance of 
  DeepEmotex models created using different batch sizes to fine-tune BERT. 
  As it can be seen in Figure \ref{bert-result} the model achieved a  performance of $91.8\%\pm0.2\%$ 
  on our test data in terms of F1 score. 
  The highest performance 92\% is achieved when the batch size is 100.  
 




\subsection{Evaluating DeepEmotex Models}
After having fine-tuned the models of  DeepEmotex, we now set out to evaluate its generality to classify  emotion. For this, 
we evaluate the performance of the DeepEmotex models to classify emotion in popular   benchmark data sets.  
An issue with many of the benchmark datasets is data scarcity, which is specially problematic
in emotion analysis \cite{felbo2017using}. 

To evaluate our DeepEmotex method on emotion detection 
we  make use of EmoInt \cite {MohammadB17wassa} and Stimulus \cite{ghazi2015detecting} as benchmark datasets. 
EmoInt (Emotion Intensity in Tweets) is a dataset of emotions in tweets from WASSA 2017 \cite{MohammadB17wassa}. The benchmark dataset provides four emotion classes including joy, anger, fear, and sadness. 
Emotion Stimulus  dataset \cite{ghazi2015detecting} contains 820 sentences annotated with both an emotion cause and an emotion tag. 
Their taxonomy is joy, anger, sadness, disgust, fear, shame and surprise. 

We evaluate three emotion classes, including joy, anger, and sadness. 
Table \ref{benchmark} lists the  number of samples in each class in  our benchmark datasets. 
As can be seen in Table \ref{benchmark},  the benchmark data sets are balanced, that is, there are approximately an equal number of examples that belong to each of the four classes. 

\begin{table}%
\caption{Number of messages in each emotion class in our  benchmark datasets}
\centering
{
\begin{tabular}{|l||p{1cm}|p{1cm}|p{1.1cm}|p{1cm}|p{1cm}|}
\hline  
Emotion Dataset &  Joy &  Sadness & Anger \\ \hline \hline
Stimulus Dataset&  480  & 540 & 570 \\ \hline
EmoInt Dataset&  671  & 642 & 656 \\ \hline

\end{tabular}
} 
\label{benchmark}
\end{table}

Our fine-tuned models are utilized to classify  the benchmark dataset. These  models classify the input benchmark dataset into our four emotion classes. 
The input class joy is correctly classified if it is labeled as happy-active or happy-inactive. The input class anger is correctly classified if it is labeled as unhappy-active, and the input class sadness is correctly classified if it is labeled as unhappy-inactive. 


To further evaluate the predictive performance of DeepEmotex models, 
we compare the classification results of our  proposed models against  the results of a baseline   model. 
We train a bidirectional LSTM as a baseline model, which have shown success in various other domains. 

Tables \ref{eval1} and \ref{eval2} list the evaluation results of the proposed DeepEmotex models to classify emotion in EmoInt dataset \cite{MohammadB17wassa} and Stimulus   dataset \cite{ghazi2015detecting} in terms of F-score. 
The first Three columns include the result of DeepEmotex-BERT and DeepEmotex-USE with freezing and unfreezing  layers. 
The last  column of the tables  lists the results of the baseline model. 
As table \ref{eval2} and \ref{eval2} presents,  
The DeepEmotex-BERT  model outperforms 
other models. 
The DeepEmotex-USE model created with unfreezing layers achieves  slightly higher classification accuracy 
than the model created with freezing layers. 

\begin{table}%
\caption{Classification accuracy of different models to predict emotion in EmoInt benchmark dataset in terms of F-score}
\centering
\resizebox{\columnwidth}{!}%
{
\begin{tabular}{|p{1.7cm}|p{1.5cm}|p{1.76cm}|p{1.5cm}|p{0.75cm}|p{0.75cm}|p{0.75cm}|}
\hline \backslashbox[21mm]{Emotion \\Class}{\\Model}
&DeepEmotex-BERT & DeepEmotex-USE UnFrozen layers & DeepEmotex-USE Frozen layers &Bi-LSTM  \\\hline\hline 
Joy&  \bf{78.4}\% & 60\% & 66\%  & 60\%  \\[6pt] \hline     
Anger& \bf{57.3}\% & 56\% & 53.3\%  & 33\% \\ [6pt] \hline      
Sadness& \bf{71.4}\% & 51.3\% & 50.4\% & 47\% \\ [6pt]\hline\hline 
Micro Average  F-score& \bf{71}\% & 58\% & 57 \% & 50\% \\ [6pt]\hline 
\end{tabular}
}
\label{eval1}
\end{table}

\begin{table}%
\caption{Classification accuracy of different models to predict emotion in Stimulus benchmark dataset in terms of F-score}
\centering
\resizebox{\columnwidth}{!}%
{
\begin{tabular}{|p{1.7cm}|p{1.5cm}|p{1.76cm}|p{1.5cm}|p{0.75cm}|p{0.75cm}|p{0.75cm}|}
\hline \backslashbox[21mm]{Emotion \\Class}{\\Model}
&DeepEmotex-BERT & DeepEmotex-USE UnFrozen layers & DeepEmotex-USE Frozen layers  &Bi-LSTM  \\\hline\hline 
Joy&  \bf{63}\% & 57\% & 60\% &  50.3\%  \\[6pt] \hline     
Anger& \bf{77}\% & 62\% & 60\% & 35\% \\ [6pt] \hline      
Sadness& \bf{77}\% & 61\% & 58.3\% & 56\% \\ [6pt]\hline\hline 
Micro Average  F-score& \bf{73.5}\% & 60\% & 59 \% & 50\% \\ [6pt]\hline 
\end{tabular}
}
\label{eval2}
\end{table}

Comparing the evaluation results of Table \ref{eval1} with Table \ref{eval2} shows that the DeepEmotex-BERT models achieve a higher performance than DeepEmotex-USE models in classifying emotion in both   benchmark datasets. 
These evaluation results show that the proposed DeepEmotex-BERT and DeepEmotex-USE models outperform 
the baseline  model (BiLSTM). 
Overall, our DeepEmotex-BERT model achieves   a higher accuracy than the baseline model (i.e., BiLSTM) by 21\% on EmoInt dataset and by 23.5\% on Stimulus dataset. 

\begin{figure}[t]
  \centering
  \includegraphics[width=\linewidth]{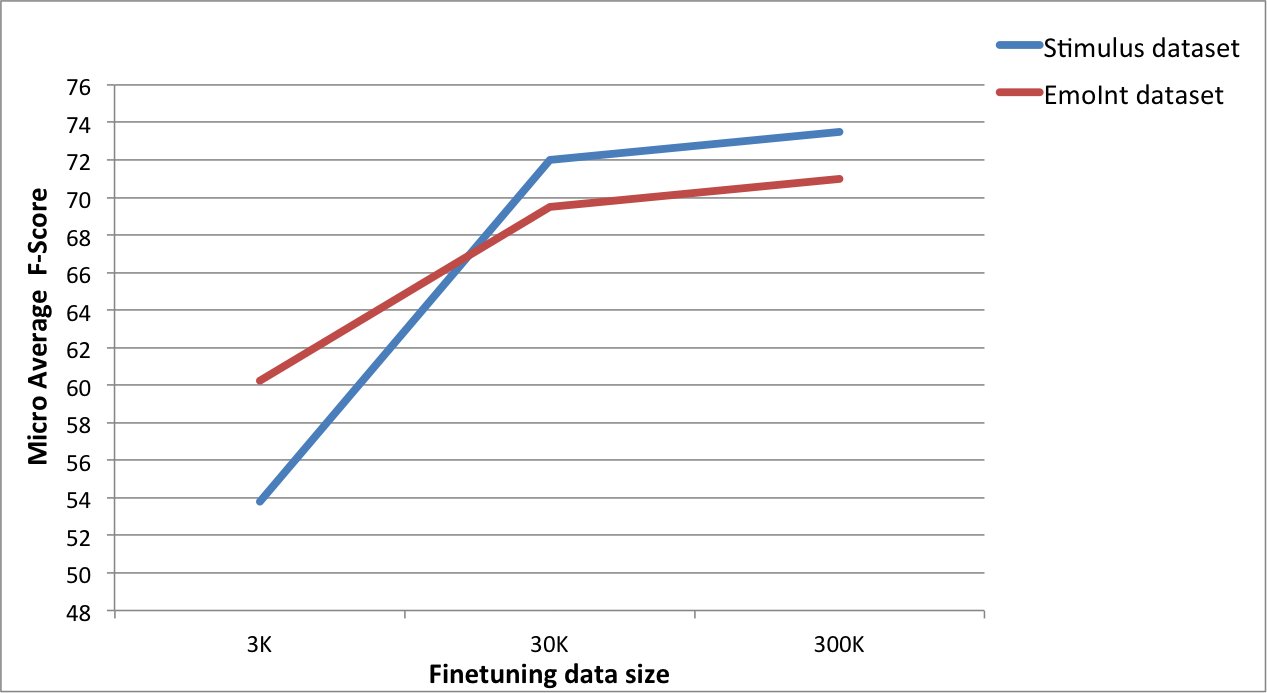}
  \caption{Effect of the size of the fine-tuning dataset on classification performance of DeepEmotx-BERT model to predict emotion of the benchmark datasets in terms of Micro Average F-score.}
  \label{finetune-size}
\end{figure}

We also investigate if the size of data during the fine-tuning task has an influence on the performance of our models.  We verify whether using more data would diminish the catastrophic forgetting problem and if it would be worthwhile gathering more  data for fine-tuning the pretrained models. Figure \ref{finetune-size} presents the  classification results of the DeepEmotex-BERT model using different size of fine-tuning dataset to predict emotion in the benchmark datasets. 
As Figure \ref{finetune-size} shows, increasing the amount of data,
with which our DeepEmotex-BERT model was fine-tuned from 3K to 30K, improves the performance of the model to classify emotion by 18\%  in the Stimulus dataset and by 9\% in the EmoInt dataset. 
Increasing the amount of fine-tuning data
from 30K to 300K slightly increased the  classification performance of our DeepEmotex-BERT model.  
The trend suggests that the proposed model is expressive enough to learn from more
data. Moreover, the proposed model reduces the effect of the forgetting problem in transferring knowledge from the pre-tranied language domain  (i.e., BERT) to our target domain.  

\section{Conclusion}
\label{conclude}
In this paper, we study  the effectiveness of deep transfer learning methods  to detect  emotion in text messages. 
We develop and evaluate DeepEmotex models to automatically classify  emotion  in  text messages. 
DeepEmotex models  learn  feature representations for emotion classification    
using sequential transfer learning.  
More precisely, DeepEmotex uses USE \cite{cer2018universal} and 
BERT \cite{devlin2018bert} as pre-trained models. 
These models are then fine-tuned on our emotion classification task and obtain   
state-of-the-art results.  

Our experimental results 
demonstrate the effectiveness and robustness of the fine-tuned 
 DeepEmotex models.  
Our DeepEmotex models  were able to achieve over  91\%  accuracy for multi-class emotion classification on test dataset. 


We evaluated the performance of the models created using BERT and the models created using USE in classifying emotion in the EmoInt benchmark dataset. 
Our DeepEmotex-BERT models were able to correctly classify emotion in 73\% of the instances in the benchmark dataset. 
Our evaluation results show that the proposed DeepEmotex-BERT models outperform 
the baseline Bi-LSTM model \cite{zhang2017ynu} by 23\%. 

We also studied the effect of varying the amount of data used for fine-tuning the pretrained models.  
We observed that using more  data for  fine-tuning  the pre-trained model improved the  classification performance of the target task by 20\%  
in the Stimulus dataset and by 11\% in the EmoInt dataset. 
\ifCLASSOPTIONcaptionsoff
  \newpage
\fi



\bibliographystyle{IEEEtran}
\bibliography{IEEEabrv,IEEEexample}
%

\end{document}